\documentstyle{aipproc}

\begin{document}

\title{Quantum algebraic representation\\
of localization and motion\\
of a Dirac electron}
\author{Marc-Thierry Jaekel $^\dagger$ and Serge Reynaud $^\ddagger$}
\address{$^\dagger$ Laboratoire de Physique Th\'eorique 
\thanks{Laboratoire du CNRS, de l'Ecole Normale Sup\'erieure et de
l'Universit\'e Paris Sud}, 
ENS, 24 rue Lhomond, F75231 Paris France\\
$^\ddagger$ Laboratoire Kastler Brossel
\thanks{Laboratoire de l'Ecole Normale Sup\'erieure,
de l'Universit\'e Pierre et Marie Curie et du CNRS},
UPMC, case 74, Campus Jussieu, F75252 Paris France}
\date{Frontier Tests of QED and Physics of the Vacuum,
Trieste, October 5-11, 2000}

\maketitle

\begin{abstract}
Quantum algebraic observables representing localization
in space-time of a Dirac electron are defined.
Inertial motion of the electron is represented in the 
quantum algebra with electron mass acting as the generator of motion.
Since transformations to uniformly accelerated frames are naturally
included in this conformally invariant description, the quantum algebra
is also able to deal with uniformly accelerated motion.
\end{abstract}

\section*{Introduction}

Modern discussions of time and space revive a basic distinction already
stated by Newton \cite{Newton}. Theoretical physics deals with two different
definitions of time and space. On one hand, the equations of motion are
written in terms of coordinate parameters representing ideal mathematical
points on a classical map of space-time. This was true for classical
Newtonian physics, but this is still true for quantum field theory or general
relativity. On the other hand, time and space measurements necessarily
reflect the properties of physical observables.

Einstein emphasized this point in his first paper on relativity
\cite{Einstein05}~:

\begin{quote}
If we wish to describe the {\it motion} of a material point, we give the
values of its co-ordinates as functions of time. Now we must bear carefully
in mind that a mathematical description of this kind has no physical meaning
unless we are quite clear as to what we understand by ``time''...

The ``time'' of an event is that which is given simultaneously with the
event by a stationary clock located at the place of the event...
\end{quote}

Einstein then explains that time indications delivered by remote clocks have 
to be compared through synchronization procedures consisting in particular
in the transfer of light pulses between the two clocks. The position of
an event in space-time can be deduced from the exchange of several light 
pulses. As is well known, these thought experiments about localization
in space-time entail that space and time are relativistic observables 
mixed under frame transformations. And they are not only thought experiments 
since they are nowadays implemented as practical applications such as
the Global Positioning System \cite{GPS91}. 

At the same time, space and time belong to the quantum domain with their
metrological definition now rooted in atomic physics \cite{Quinn91}. 
The time delivered by an atomic clock is the phase of a quantum oscillator 
while electromagnetic signals used to synchronize remote clocks are quantum 
fields. This raises the question of the compatibility of quantum and 
relativitistic descriptions of space and time, as explicitly stated by 
Schr\"{o}dinger \cite{Schrodinger30}~:

\begin{quote}
In Lorentz transformations, time and space coordinates enter in a completely
symmetrical manner. But in quantum mechanics, ... time is a quite different
thing than the co-ordinates.

Time is not treated as an observable... It is a parameter the value of which
is supposed to be exactly known~: it is in fact the old good time of Newton
and quantum mechanics does not worry about the existence of the old good
clock which it would need in order to know the value of this parameter {\it t}.

But it seems to me doubtless that we will have to give up this too classical
notion of time, and not only because of relativity...

The knowledge of the variable {\it t} is obtained in the same manner as that
of any other variable, by observing a physical system, namely a clock. {\it t%
} is therefore an observable and must be treated as an observable~; time
must in general have ``statistics'' and not a ``value''.
\end{quote}

In other words, space and time have to be treated as quantum observables
with their proper quantum fluctuations. And, moreover, this treatment has to
be compatible with the effects of relativistic frame transformations. It is
well known in ordinary quantum mechanics that space positions may be
represented as quantum operators conjugated to momentum operators. This is
still the case in standard quantum field theory \cite{Pryce48,NewtonW49}.
But it is also commonly thought that standard quantum formalism does not
allow for time being treated as an operator conjugated to energy \cite
{Pauli58,Wightman62} which forbids to relate the energy-time Heisenberg
inequality to a quantum commutation relation. This also entails 
that Lorentz transformations cannot be represented properly with positions 
in space described by quantum observables 
and position in time described by a classical coordinate parameter.

Lorentz invariance may be restored by abandoning the
observable character of space variables as well \cite{Heisenberg30}. This
clears up the unacceptable difference between space and time with however
the drawback of no longer having explicit representations of space-time
observables. Physical fluctuations are then described by quantum fields. 
This situation makes the quantum implementation of relativistic symmetries 
quite unsatisfactory \cite{deWitt63} and plagues the attempts to construct 
a quantum theory including gravity \cite{Rovelli91,Isham97}. Coming back
to the quotation of Einstein, this also challenges the very representation
of ``motion'' in quantum theory.

\section*{Localization observables}

We develop here an alternative approach to localization and motion in 
space-time where quantum and relativistic requirements are treated 
simultaneously and consistently \cite{QED98,EPJD99}. 
The main idea is to define localization observables generalizing 
Einstein's definition of the position of the center-of-mass in special 
theory of relativity \cite{Einstein06}. The latter definition relies on 
the invariance of equations of motion under Lorentz frame transformations. 
Positions in space are then expressed in terms of the symmetry generators, 
{\it i.e.} the components $P_{\mu }$ of energy-momentum vector and the 
components $J_{\mu \nu }$ of angular momentum tensor. 
The more general definition advocated here uses invariance of Maxwell equations 
not only under Lorentz frame transformations but also under the larger group of
conformal transformations. This definition will allow one to give a quantum
description of motion of an electron, as it was demanded by Schr\"{o}dinger.

Conformal symmetry of electromagnetism has been discussed by Bateman and 
Cunningham \cite{Bateman09,Cunningham09}, very early after the
advent of relativity theory.  
It has been studied in a number of papers \cite{Fulton62} and it still holds for
free electromagnetic fields in quantum field theory \cite{Binegar83}. In
particular conformal symmetry includes a dilatation generator $D$ which,
together with the Poincar\'{e} generators, allows one to construct 4 quantum
observables $X_{\mu }$ representing positions in space-time. We recall below
the main building blocks of this construction described in more details
in \cite{QED98,EPJD99}. 

The whole framework is based upon the conformal algebra, that is the set of
commutators which characterize conformal symmetry. Conformal algebra firstly
contains a sub-algebra with Poincar\'{e} and dilatation generators  
\begin{eqnarray}
&&\left( P_{\mu },P_{\nu }\right) =0\qquad \left( J_{\mu \nu },P_{\rho
}\right) =\eta _{\nu \rho }P_{\mu }-\eta _{\mu \rho }P_{\nu }  \nonumber \\
&&\left( J_{\mu \nu },J_{\rho \sigma }\right) =\eta _{\nu \rho }J_{\mu
\sigma }+\eta _{\mu \sigma }J_{\nu \rho }-\eta _{\mu \rho }J_{\nu \sigma
}-\eta _{\nu \sigma }J_{\mu \rho }  \label{PJD} \\
&&\left( D,P_{\mu }\right) =P_{\mu }\qquad \left( D,J_{\mu \nu }\right) =0 
\nonumber
\end{eqnarray}
$P_{\mu }$ and $J_{\mu \nu }$ are the translation and rotation generators,
that is also components of energy-momentum vector and angular momentum
tensor. $D$ is the dilatation generator producing a change of space-time and
energy-momentum units which preserves the velocity of light $c$ and the
Planck constant $\hbar $. Quantum algebraic commutators (\ref{PJD})
represent the relativistic shifts of observables under translations and
rotations with $\eta _{\mu \nu }={\rm diag}\left( 1,-1,-1,-1\right) $
standing for Minkowski tensor as well as the commutation relations of
observables. We use the following notations for the commutator, the
symmetrized product and the symmetrized division of observables 
\begin{equation}
\left( A,B\right) \equiv \frac{AB-BA}{i\hbar }\qquad A\cdot B\equiv \frac{%
AB+BA}{2}\qquad \frac{A}{B}\equiv A\cdot \frac{1}{B}
\end{equation}

Localization observables are constructed from the algebra (\ref{PJD}) as the
solutions $X_{\mu }$ of the equations 
\begin{equation}
J_{\mu \nu }=P_{\mu }\cdot X_{\nu }-P_{\nu }\cdot X_{\mu }+S_{\mu \nu
}\qquad D=P^{\mu }\cdot X_{\mu }  \label{defX}
\end{equation}
Angular momenta $J_{\mu \nu }$ are sums of orbital and spin contributions.
As the spin tensor $S_{\mu \nu }$ is transverse with respect to momentum,
the first equation fixes the transverse components of position observables
with respect to momentum. This explains why $D$ has to be involved to fix
the longitudinal components of position observables. Position observables
are obtained as the following expressions which generalize Einstein's
definition of the center-of-mass position as soon as the square 
$P^{2}\equiv P_{\mu }P^{\mu }$ of the momentum vector differs from 0 
\begin{equation}
X_{\mu }=\frac{D\cdot P_{\mu }-J_{\mu \nu }\cdot P^{\nu }}{P^{2}}
\label{solX}
\end{equation}
Elementary algebraic calculus then leads to their
transformation properties 
\begin{eqnarray}
\left( P_{\mu },X_{\nu }\right)  &=&-\eta _{\mu \nu }  \nonumber \\
\left( J_{\mu \nu },X_{\rho }\right)  &=&\eta _{\nu \rho }X_{\mu }-\eta
_{\mu \rho }X_{\nu }  \label{PX} \\
\left( D,X_{\mu }\right)  &=&-X_{\mu }  \nonumber
\end{eqnarray}
These equations respectively mean that position observables are canonically
conjugated to momenta, that they are transformed as components of a Lorentz
vector under Lorentz transformations and that they have a conformal weight
opposite to that of momenta. These results meet the expectations
based on classical relativity but they bear on quantum observables. 

However different position components are found to have a non null
commutator which is directly connected to spin, implying that 
dispersions obey a Heisenberg inequality 
\begin{equation}
\left( X_{\mu },X_{\nu }\right) =\frac{S_{\mu \nu }}{P^{2}}  \qquad
\left( \Delta X\right) ^{2} \gtrsim \frac{\hbar ^{2}}{P^{2}}  \label{XX}
\end{equation}
This important output of the quantum algebraic formalism clearly indicates
that the conceptions of space-time inherited from classical relativity have
to be revised for quantum objects. Positions in space-time cannot be associated
with sizeless classical points but they have rather to be thought of as fuzzy
spots with a size $\Delta X$ given by Compton relation. 

It is possible to give a geometrical interpretation of these results by
coming back to the Einstein construction of a position in space-time as the
intersection of 2 light rays propagating in different directions. For such a
field state, the squared mass $P^{2}$ associated with the field state
differs from 0. It may then be proved that positions (\ref{solX}) correspond
to the coincidence of the two light rays. Precisely, two real light rays never
intersect exactly and each ray has a transverse dimension
due to diffraction but the positions correspond to the middle point on the
segment which joins the two rays while being perpendicular to both rays \cite
{EPJD99}. The fuzziness in (\ref{XX}) due to the spin associated with the 
2-light-rays state is directly connected to the length of the segment involved 
in the geometrical construction.

Let us recall at this point that conformal symmetry includes four additional
generators $C_{\mu }$ which represent transformations to uniformly accelerated 
frames. The algebra (\ref{PJD}) is thus complemented by the commutators 
\begin{eqnarray}
&&\left( D,C_{\mu }\right) =-C_{\mu }\qquad \left( C_{\mu },C_{\nu }\right)
=0  \nonumber \\
&&\left( J_{\mu \nu },C_{\rho }\right) =\eta _{\nu \rho }C_{\mu }-\eta _{\mu
\rho }C_{\nu } \\
&&\left( P_{\mu },C_{\nu }\right) =-2\eta _{\mu \nu }D-2J_{\mu \nu } 
\nonumber
\end{eqnarray}
Generators $C_{\mu }$ are commuting components of a Lorentz vector with a
conformal weight opposite to that of momenta. Their commutators $\left(
P_{\mu },C_{\nu }\right) $ with translations describe the redshifts of
momenta under transformations to accelerated frames and thus constitute
quantum versions of the Einstein redshift law \cite{EPL97}. This entails
that the quantum algebraic construction allows one to discuss relativistic
effects associated with accelerated frames from invariance properties
represented by conformal algebra \cite{FoP98}. We will show below that the
same algebraic framework allows one to represent inertial as well as
uniformly accelerated motion.

\section*{Conformal Dirac electron}

Up to now, localization observables have been constructed on Einstein
procedures with light rays so that conformal symmetry has been used in the
case of free electromagnetic fields. The localization
observables thus constitute a theoretical representation equivalent to
standard QED theory although the interpretation may be quite different from
the standard one. In particular conformal symmetry has been used for the
field state associated with 2-light-rays localization although the
mass does not vanish for this state. Also, conformal generators are
space-time integrals of the quantum stress tensor and localization
observables are highly non linear and non local expressions of the fields.
Since they are not defined in vacuum or one photon states, these hermitian
observables are not self-adjoint. This is just the point where the Pauli
theorem forbidding the definition of a quantum time operator has been
circumvented \cite{QED98,EPJD99}. 

We want now to show that quantum algebra also allows one to deal with Dirac
electrons \cite{PLA99}. To this aim, we first introduce a second set $x_{\mu
}$ of position observables and a new spin tensor $s_{\mu \nu }$.
Poincar\'{e} and dilatation generators keep the same form (\ref{defX}) when 
$X_{\mu }$ and $S_{\mu \nu }$ are replaced by $x_{\mu }$ and $s_{\mu \nu }$ but
the new observables obey canonical commutation relations. In particular the
components $x_{\mu}$ commute with other components of positions $x_{\nu}$
or spin $s_{\nu \rho}$. The canonical positions may be written 
\begin{equation}
x_{\mu }=X_{\mu }-i\gamma \frac{W_{\mu }}{P^{2}}\qquad W^{\mu }\equiv -\frac{%
1}{2}\epsilon ^{\mu \nu \rho \sigma }J_{\nu \rho }P_{\sigma }  
\qquad \gamma ^{2}=1  \label{solx}
\end{equation}
where $W^{\mu }$ is the Pauli-Lubanski spin vector and $\gamma $ is a sign 
representing the orientation of the spin tensor $s_{\mu \nu }$.
$\gamma $ is invariant under translations, rotations and dilatation and plays the
same role here as $\gamma _{5}$ in Dirac electron theory \cite{Dirac28}.

Canonical positions are transformed under translations, rotations and
dilatation as expected from classical relativity~: they obey equations (\ref
{PX}) with $X_{\mu }$ replaced by $x_{\mu }$. This entails that the
requirements enounced by Schr\"{o}dinger have now been met. Positions may be
represented as quantum relativistic observables which are conjugate with
respect to momentum observables while properly representing Lorentz
symmetry. A position in time has been defined besides positions in space and
it is conjugated to energy as positions in space are conjugated to spatial
momenta. The 4 positions in space-time are mixed under Lorentz
transformations according to the classical laws. It has however to be
emphasized that the canonical variables $x_{\mu }$ and $s_{\mu \nu }$ are
not hermitian, as shown by (\ref{solx}). This is a further important output
of the quantum algebraic approach to the localization problem. One may
define either hermitian observables with non canonical commutators 
or canonical variables which are not hermitian.

In order to build up a quantum algebraic theory of electrons, we use the
general properties already deduced from conformal symmetry and add two
further assumptions. First we assume that the spin number is $\frac{1}{2}$ 
\begin{equation}
\frac{W^{2}}{P^{2}}=-\hbar ^{2}s\left( s+1\right) \qquad s=\frac{1}{2}
\label{defs}
\end{equation}
and invariant under transformations to accelerated frames. As it is already
invariant under translations, rotations and dilatation, $s$ is thus
invariant under the whole conformal algebra. Then we introduce the sign 
$\varepsilon $ of mass which commutes with translations, rotations and 
dilatation. $\gamma $ commutes with $\varepsilon ^{2}=1$ while $\varepsilon $ 
commutes with $\gamma ^{2}=1$ and these conditions are fulfilled as soon as 
$\gamma $ and $\varepsilon $ either commute or anticommute. 
We assume here that they anticommute 
\begin{equation}
M=\varepsilon \sqrt{P^{2}}\qquad \varepsilon ^{2}=1
\qquad \gamma \cdot \varepsilon =0  \label{anticom}
\end{equation}
Hence the spin orientation $\gamma $ changes the mass sign $\varepsilon $ 
into its opposite while the mass sign $\varepsilon $ changes the spin 
orientation $\gamma $ into its opposite.

From these assumptions, we deduce that there exist Clifford symbols in the
quantum algebra which allow one to write a Dirac equation 
\begin{equation}
\gamma _{\mu }=\frac{P_{\mu }}{\hbar M}-2\gamma 
\frac{W_{\mu }}{\hbar M} \qquad 
\gamma _{\mu}\cdot \gamma _{\nu }=\eta _{\mu \nu }  \qquad
M=P^{\mu }\gamma _{\mu }=\gamma _{\mu }P^{\mu }  
\label{Dirac}
\end{equation}
The symbols $\gamma _{\mu }$ commute with canonical positions and momenta
and their commutators reproduce the canonical spin tensor $s_{\mu \nu }$.
These relations constitute a quantum algebraic extension of Dirac electron
theory \cite{Dirac28}. At this point, it is worth emphasizing that the
Clifford relations and the Dirac equation have been derived from conformal 
symmetry and the 2 further assumptions (\ref{defs},\ref{anticom}). 

The quantum algebraic formalism leads to a striking difference with the
standard Dirac theory. The mass $M$ is now a quantum operator which, like
its sign $\varepsilon $, anticommutes with $\gamma $. Though being a Lorentz
scalar invariant under translations and rotations, $M$ is not invariant under
dilatation since it has the same conformal weight as momenta. These
properties are certainly incompatible with the classical treatment of the mass
commonly associated with Dirac electron. Notice however that electron mass is,
at least partly, generated by electromagnetic self-energy and that it should
therefore present intrinsic quantum fluctuations. Since electron-positron
pairs may decay into 2-photon pairs, it seems hard to forbid treating the mass of
the $e^{-}-e^{+}$ pair similarly to that of the 2-photon pair. But, as
already noticed, the second one is commonly considered to obey conformal
invariance. 

These arguments plead for mass having its proper conformal dimension, 
so that the conformal symmetry holds for free electron states as well 
as for photon states. They also correspond to the motivations of Weyl 
aiming at a conformal description of space-time \cite{Weyl22} or Dirac
attempting to describe the electron field in a conformal space \cite
{Dirac36}. In any case, modern descriptions of the electron are no longer
identical to the original Dirac theory. Electron mass is now considered to
be generated through an interaction with Higgs fields \cite{Itzykson85} and
Higgs models obeying conformal invariance are available \cite{Pawlowski98}.
Here, we do not specify a particular field model but we consider that the
treatment of electron mass is compatible with conformal symmetry and, thus,
given by the previously written equations.

\section*{Inertial motion}

As just discussed, mass $M$ is a quantum observable. We show now that
commutators with this observable play an important role since they generate
inertial motion. To this aim, we introduce a prime symbol 
representing an algebraic derivative and obeying Leibniz rule 
\begin{equation}
F^{\prime }=\left( F,M\right) \qquad \left( FG\right) ^{\prime }=F^{\prime
}G+FG^{\prime }  \label{defPrime}
\end{equation}
The same property is obtained for the commutator with any observable but the present
definition implies that the Poincar\'{e} generators are constants of motion
$P_{\mu }^{\prime }=J_{\mu \nu }^{\prime }=0$.
These features resemble the common Hamiltonian formalism with however the
mass as the motion generator and, hence, a full
compatibility with Lorentz invariance. The derivative defined by (\ref
{defPrime}) may be thought of as associated with a quantum ``evolution time'' 
$\frac{D}{M}$ since the latter observable has a unit derivative 
\begin{equation}
D^{\prime }=M\qquad M^{\prime }=0\qquad \left( \frac{D}{M}\right) ^{\prime
}=1
\end{equation}
This evolution time is distinct from the date $X_{0}$ of an event,
that is its position in time. Meanwhile, both times have to be
distinguished from their classical analogs. 

The motion of hermitian positions corresponds to the simple expectations of
classical mechanics. The hermitian velocity, that is the derivative of the
hermitian position, is the standard mechanical velocity given by the
ratio of momentum to mass while the hermitian acceleration vanishes 
\begin{equation}
X_{\mu }^{\prime }=\frac{P_{\mu }}{M}\qquad X_{\mu }^{\prime \prime }=0
\end{equation}
Hermitian spin is also conserved. The orientation $\gamma $ 
oscillates at twice the rest mass frequency 
\begin{equation}
\gamma ^{\prime }=\frac{2}{i\hbar }\gamma M=-\frac{2}{i\hbar }M\gamma \qquad
\gamma ^{\prime \prime }=-\frac{4M^{2}}{\hbar ^{2}}\gamma 
\end{equation}
It follows that the canonical velocities are identical to the Clifford
generators and that they undergo a ``Zitterbewegung'' 
\begin{equation}
x_{\mu }^{\prime }=\gamma _{\mu }\qquad x_{\mu }^{\prime \prime }=\gamma
_{\mu }^{\prime }=-i\gamma ^{\prime \prime }\frac{W_{\mu }}{P^{2}}
\end{equation}
These relations are analogous to well known results of standard Dirac theory
with, once again, an algebraic representation of the motion generated
by mass observable. 

\section*{Accelerated motion}

As already explained, the quantum algebraic formalism relies upon conformal
symmetry and, therefore, has the ability of dealing with uniformly
accelerated frames as well as inertial frames. It can therefore 
describe uniformly accelerated motion as well as inertial motion. In order
to prove this statement, we write the shift of the mass observable from $M$
in a frame to $\overline{M}$ in another frame as a conjugation by an element
of the conformal group 
\begin{equation}
\overline{M}=\exp \left( -\frac{a^{\rho }C_{\rho }}{2i\hbar }\right) M\exp
\left( \frac{a^{\rho }C_{\rho }}{2i\hbar }\right) =M-\frac{a^{\rho }}{2}%
\left( C_{\rho },M\right) +\ldots   \label{traM}
\end{equation}
The parameters $a^{\rho }$ are classical accelerations along the $4$
space-time directions. Here, we restrict our attention to the linear
approximation with respect to these parameters. It is important to notice
that conjugations preserve the structure of quantum algebraic relations. For
example, position and momentum observables are transformed under conjugation
but the canonical commutators between them are preserved since they are
classical numbers. Canonical commutators have the same form in accelerated
and inertial frames and are written in terms of the same Minkowski metric $%
\eta _{\mu \nu }$. This result had to be expected in a quantum algebraic
approach but it clearly stands in contradistinction with covariant
techniques of classical relativity.

Relation (\ref{traM}) gives the redshift of mass under the frame transformation 
\begin{equation}
\left( C_{\rho },M\right) =2M\cdot X_{\rho }
\end{equation}
The redshift has exactly the form expected from Einstein classical law since
it is proportional to $M$ and to the gravitational potential $a^{\rho
}X_{\rho }$ arising from the transformation according to Einstein
equivalence principle. The shift may also be read as a conformal metric
factor which depends on position observables as the classical metric factor
depends on classical coordinates \cite{PLA99}. To fix ideas, we may consider
the coordinate frame denoted with bars to be an inertial frame and the coordinate
frame without bars to be the accelerated frame. We may then define ``inertial
motion'' as the algebraic derivative $F^{\prime }=\left( F,\overline{M}\right)$ 
associated with the ``inertial mass'' $\overline{M}$. This choice leads to 
conservation of Poincar\'{e} generators $\overline{P}_{\mu}$ and 
$\overline{J}_{\mu \nu }$ defined in the inertial frame. The laws
of inertial motion have also the same form as previously in the inertial
frame. Now we can write the motion of observables $P_{\mu}$ or $J_{\mu \nu }$ 
as they are defined in the accelerated frame. 
Proceeding in this manner we obtain for the hermitian positions $X_{\mu}$ 
\begin{equation}
X_{\mu }^{\prime \prime }=a_{\mu }+\ldots 
\end{equation}
This is a quantum algebraic expression of the law of free fall in the
gravity force arising in the accelerated frame from the Einstein equivalence
principle. The dots mean that this expression is linearized in the
gravity acceleration. A more general expression with non linear terms is
deduced from conformal symmetry in \cite{AdP00}. 

\section*{Conclusions}

We have shown that 
positions in space-time may be represented as quantum relativistic
observables conjugated with respect to momentum observables while
properly obeying Lorentz symmetry. The whole framework constitutes a
``quantum algebraic relativity'' where relativistic effects under frame
transformations and quantum commutation relations are described by a unique
algebraic calculus. Mass is a quantum observable and this has important
consequences. First, it suffers shifts under transformations to accelerated
frames and these shifts reproduce the effect of the gravitational potential
arising as a consequence of Einstein equivalence principle. Then, mass is
the generator of inertial motion and the choice of inertial frames among
all possible definitions allows one to represent uniformly accelerated
motion as well, still in full consistency with Einstein equivalence
principle.

The quantum algebraic description of localization in space-time relies on
conformal symmetry and it can be written so that it explicitly
displays invariance under the ${\rm SO}\left( 4,2\right) $ algebra \cite
{AdP00}. This allows one to write the laws of inertial motion as well as the
Newton equation of motion in a gravity field under an invariant form 
consistent with this algebra. The whole description may thus be thought of
as a ``quantum geometry'' \cite{Connes94} relying on conformal symmetry.

\end{document}